# General analysis of the complementary nature of coercivity enhancement and exchange bias in ferro-antiferromagnet (F-AF) exchange coupled systems

Congxiao Liu and Min Sun

Department of Mathematics and Center for Materials for Information Technology,

University of Alabama, Tuscaloosa, Alabama 35487-0350, USA

Hideo Fujiwara

Department of Physics and Astronomy and Center for Materials for Information

Technology, University of Alabama, Tuscaloosa, Alabama 35487-0209, USA

Complementary nature of coercivity enhancement and exchange bias is generalized from the layered systems to ferro-antiferromagnet (F-AF) exchange coupled systems with arbitrary configurations and is proved based on the coherent rotation of the F magnetization. In the proof, the effect of F-AF coupling is absorbed into the anisotropy of the F part, resulting in an arbitrary anisotropy for the F part. The proof starts from a general discussion on the initial susceptibility of a single domain particle. Then the fundamental correlation between the maximal initial susceptibility and the critical field along an arbitrary easy direction of a single domain particle, at which the magnetization becomes unstable, is discussed. Finally, the difference of the initial switching field and the actual switching field along the arbitrarily chosen easy direction is discussed.

## I. Introduction

Exchange anisotropy was discovered by Meiklejohn and Bean[1] in 1956. It originates from the exchange interactions between the spins of the ferromagnet (F) and the spins of the antiferromagnet (AF) at their interface. It is a complicated phenomenon that depends on a variety of factors including the ferro-antiferromagnet (F-AF) interface nature and the properties of the antiferromagnet. The difficulty of monitoring the F-AF interface and the lack of knowledge about the properties of antiferromagnets make the study of this anisotropy rather difficult. Furthermore, because of the F-AF coupling, the AF spins are not rigid during F magnetization processes. As a result, the effective anisotropy of the F part due to the F-AF coupling is not easy to predict. In fact, study of the $Ni_{80}Fe_{20}$/FeMn F-AF bilayers showed that the angular dependences of the exchange bias field and the coercivity of the $Ni_{80}Fe_{20}$ were sensitive to the film growth direction[2] and were different for epitaxial films and textured films[3].

Despite these difficulties, there are some general features for the F-AF systems that are independent of the details of the systems. For example, it was found that when a ferromagnet is coupled to an antiferromagnet through their interface, its magnetization behavior is greatly altered; namely, the magnetization tends to be more uniform upon reversal compared with the uncoupled single ferromagnet, a law first found by Fujiwara et al[4]. This fact is evidenced by the experimental results of the F-AF exchange bilayers from the AMR measurements[4], MFM observations[5], torque curves[6], angular dependences of the exchange bias field and the coercivity of the pinned layer[7], etc. But the most important experimental result showing the coherent rotation of the F magnetization is the so-called complementary nature of coercivity enhancement and exchange bias[4 8 9 10], which was proposed by Fujiwara[4]. For an F-AF exchange bilayer, one can define the intrinsic pinning field $H_{p0}$ by $M_s/\chi_{in\perp}$, where $M_s$ is the saturation magnetization of the F layer and $\chi_{in\perp}$ is the initial susceptibility transverse to the stable direction of the F magnetization in the absence of external field. This definition is a natural generalization of the anisotropy field for a single F layer. It was found that $H_{p0}$ is approximately equal to the switching field of the F magnetization along the stable direction of the F magnetization in the absence of external field, denoted by $H_{sw}^-$, which is the sum of the total coercivity $H_c$ of the F layer and the exchange bias field $H_{eb}$. This reveals that in the F-AF bilayers both the coercivity enhancement of the F layer and the exchange bias originate from the F-AF coupling, and that the magnetization reversal process of the F layer is dominated by coherent rotation. Theoretically it was shown that[8], in magnitude, $H_{p0}$ is precisely equal to the F layer's *initial* switching field along the referred stable direction $H_{sw0}^-$, defined as the reverse field at which the F magnetization starts to be unstable. The difference between the actual switching field and the initial switching field will be discussed in more detail in section IV. To avoid confusion, from now on, we use the term of the *complementary nature of coercivity enhancement and exchange bias* in the F-AF bilayers to refer to the exact equivalence relationship between $H_{p0}$ and $H_{sw0}^-$. The original mathematical proof assumed the magnetization to rotate in plane[8]. However, the key to our proof was only the coherence of the F magnetization.

Exchange anisotropy was first found in a compact of fine particles of cobalt with a cobaltous oxide shell[1], for which the F-AF coupling occurs at the Co-CoO interface. By nature it was not a layered system. Although the complementary nature was found in F-AF bilayers, one would expect a similar relationship in a general F-AF system with arbitrary configuration, such as the Co-CoO system and a trilayer AF-F-AF system. However, the generality of spin configuration makes it necessary to redefine $H_{p0}$. For a general F-AF system, we may define $H_{p0}$ as $M_s/\chi_{in,max}$, where $\chi_{in,max}$ is the maximum initial susceptibility (over all the directions in the 3-d space) attained in some direction perpendicular to the stable direction of the magnetization in the absence of external field. It is the purpose of this paper to prove, based on the coherence of the F magnetization, $H_{p0} = H_{sw0^-}$ for a general F-AF system, with the definition of $H_{sw0^-}$ being the same as for the layered systems. The strategy for the proof of this particular result is that the effect of the F-AF coupling can be absorbed into the anisotropy energy of the F part. That is, the overall anisotropy of the F part includes its intrinsic anisotropy as well as the anisotropy induced by the F-AF coupling. By doing that the complication of microscopic F-AF interactions at the interface can be technically avoided and the generality of the discussion is achieved. In fact the same approach allows us to discuss any type of F-AF system. Note that the F-AF coupling induced anisotropy is not limited to a unidirectional anisotropy plus a uniaxial anisotropy[2]. In order for our discussion to be valid for any ferromagnetic materials for the F part of system, we do not assume any specific form for the intrinsic anisotropy of the F part, either. With these factors being taken into account, the overall anisotropy of the F part in our proof will be of arbitrary form. The proof for the complementary nature of the coercivity enhancement and exchange bias in F-AF coupled systems thus becomes a problem of an uncoupled single domain particle with arbitrary anisotropy.

For a general 3-d F-AF system for which the initial state is given, the magnetic state includes the AF domain structure and is path dependent due to the switching of the AF grains during the F magnetization processes. The overall anisotropy of the F part does not exist for the whole state space. However, in the initial susceptibility measurement the F magnetization only undergoes small deviation from the initial stable direction. Thus if the physical processes for perturbations of the F magnetization around the initial stable direction are reversible, the overall anisotropy is well defined in the neighborhood of the initial stable direction. For a 2-d F-AF system, that constraint can be relaxed because the F magnetization rotates in plane. During its switching process it has only one path to undergo and always rotates in one sense. Therefore for a 2-d F-AF system, although hysteresis occurs, the overall anisotropy is well defined except at switching of the F magnetization, and it is thought to be continuous during the continuous rotation of the F magnetization. Thus the overall anisotropy is well defined in the initial susceptibility measurement for a 2-d system. For the following discussions to be valid, we require that the second derivatives of the overall anisotropy exist in a neighborhood of the initial F magnetization direction.

In the following we start by a general discussion of the initial susceptibility for a single domain particle in section II. In section III we show $H_{p0} = H_{sw0^-}$. In section IV we discuss the difference between the initial switching field and the actual switching field. We conclude in section V.

## II.     Initial susceptibility

For any single domain particle, the direction of the magnetization **M** determines the magnetic state of the particle. Without loss of generality, we assume **M** initially points to the positive *x* direction, one of the easy directions of the particle. In an F-AF system, that corresponds to one of the stable directions in the absence of external field, which could be different from the pinned direction (see section IV). At this stage, the other two coordinate axes can be specified arbitrarily to form a standard 3-d rectangular coordinate system. Suppose that an external field **H** is then applied and the magnetization rotates to a new direction. Spherical coordinates ($\theta$, $\varphi$) and ($\alpha$, $\beta$) are used to specify the directions of **H** and **M**, respectively. Fig. 1 sketches the directions of **H** and **M**.

Let $E(\alpha,\beta)$ be the anisotropy energy density of the particle. Denote $\partial E/\partial \alpha$ and $\partial^2 E/\partial \alpha^2$, the first two partial derivatives of $E(\alpha,\beta)$ with respect to $\alpha$, by $E_\alpha$ and $E_{\alpha\alpha}$, etc. The assumption for an easy direction at the positive x direction (i.e., $\alpha = \pi/2, \beta = 0$) ensures the existence of the partial derivatives of $E(\alpha,\beta)$ at $\alpha = \pi/2, \beta = 0$. Furthermore, we have $E_\alpha(\pi/2,0) = 0$, $E_\beta(\pi/2,0) = 0$, $E_{\alpha\alpha}(\pi/2,0) > 0$, $E_{\beta\beta}(\pi/2,0) > 0$ and $E_{\alpha\alpha}(\pi/2,0)E_{\beta\beta}(\pi/2,0) > [E_{\alpha\beta}(\pi/2,0)]^2$. For simplicity, we can properly select y-axis and z-axis so that $E_{\alpha\beta}(\pi/2,0) = 0$ (the proof for that is in appendix A).

The free energy density $F$ of the particle includes the anisotropy energy and Zeeman energy

$$F(\alpha,\beta) = E(\alpha,\beta) - M_s H[\sin\theta\sin\alpha\cos(\beta-\varphi) + \cos\theta\cos\alpha], \tag{1}$$

where $M_s$ is the saturation magnetization.

At any stage of initial susceptibility measurement, the first derivatives of the free energy with respect to α and β vanish, yielding

$$E_\alpha - M_s H[\sin\theta\cos\alpha\cos(\beta-\varphi) - \cos\theta\sin\alpha] = 0, \tag{2}$$

$$E_\beta + M_s H \sin\theta\sin\alpha\sin(\beta-\varphi) = 0. \tag{3}$$

To calculate susceptibility, we take a variation $\delta\mathbf{H}$ of the external field along its original direction. This results in a change of the magnetization direction, denoted by $\delta\alpha$ and $\delta\beta$. Thus the new direction of the magnetization is $\alpha = \delta\alpha + \pi/2$, $\beta = \delta\beta$. From the variations of (2) and (3), setting $\alpha = \pi/2$, $\beta = 0$ and $H = 0$ for initial susceptibility, we can solve the resulting variational equations for $\delta\alpha/\delta H$ and $\delta\beta/\delta H$ to obtain

$$\delta\alpha / \delta H = -M_s \cos\theta / E_{\alpha\alpha}(\pi/2, 0),$$

$$\delta\beta / \delta H = M_s \sin\theta \sin\varphi / E_{\beta\beta}(\pi/2, 0).$$

From the definition of the initial susceptibility $\chi_{in}$ we obtain

$$\chi_{in} \equiv \delta M_{//} / \delta H$$

$$= M_s[-\cos\theta \, \delta\alpha / \delta H + \sin\theta \sin\varphi \, \delta\beta / \delta H]$$

$$= \frac{M_s^2 [E_{\alpha\alpha}(\pi/2, 0) \sin^2\theta \sin^2\varphi + E_{\beta\beta}(\pi/2, 0) \cos^2\theta]}{E_{\alpha\alpha}(\pi/2, 0) E_{\beta\beta}(\pi/2, 0)}$$

$$= \frac{M_s^2 [E_{\alpha\alpha}(\pi/2, 0) e^2_{Hy} + E_{\beta\beta}(\pi/2, 0) e^2_{Hz}]}{E_{\alpha\alpha}(\pi/2, 0) E_{\beta\beta}(\pi/2, 0)}, \tag{4}$$

where $e_{Hy} = \sin\theta \sin\varphi$ and $e_{Hz} = \cos\theta$ are the direction cosines of the external field **H**, and

$$M_{//} = M_s[\sin\theta \sin\alpha \cos(\beta - \varphi) + \cos\theta \cos\alpha]$$

is the parallel component of the magnetization along **H**.

From (4), we see that $\chi_{in}$ has a great deal of symmetries. Namely, it is symmetric with respect to the *x-y*, *y-z*, and *x-z* planes. This symmetry can be easily verified from the definitions of $e_{Hy}$ and $e_{Hz}$. Furthermore, $\chi_{in}$ depends on $E(\alpha,\beta)$ only through $E_{\alpha\alpha}(\pi/2,0)$ and $E_{\beta\beta}(\pi/2,0)$.

The above properties of $\chi_{in}$ are reasonable since $\chi_{in}$ involves the response of the magnetization to the perturbation of an infinitesimal field. Thus it depends only on the local anisotropy as the magnetic state is being perturbed. The symmetry of the local anisotropy determines the symmetry of $\chi_{in}$. In our case, the perturbed state is around $\alpha = \pi/2$, $\beta = 0$. Near this direction, the anisotropy energy density $E$ at $\alpha = \pi/2 + \delta\alpha$, $\beta = \delta\beta$ is (to the second order)

$$E(\alpha,\beta) = E(\pi/2,0) + [E_{\alpha\alpha}(\pi/2,0)(\delta\alpha)^2 + E_{\beta\beta}(\pi/2,0)(\delta\beta)^2]/2. \qquad (5)$$

Note that the first derivatives $E_\alpha$ and $E_\beta$ vanish at $\alpha = \pi/2$, $\beta = 0$ since it is an easy direction, and the coordinate system has been chosen properly so that $E_{\alpha\beta}(\pi/2,0) = 0$. From (5) it is clear that $E(\alpha,\beta)$ possesses a high degree of symmetry and is completely determined by $E_{\alpha\alpha}(\pi/2,0)$ and $E_{\beta\beta}(\pi/2,0)$. This is consistent with the properties of $\chi_{in}$. A 3-d plot of $\chi_{in}$ with $E_{\alpha\alpha}(\pi/2,0) / E_{\beta\beta}(\pi/2,0) = 1/2$ is shown in Fig. 2 (where radii represent the values of $\chi_{in}$).

# III. Initial switching field $H_{sw0}^-$ along an easy direction and the correlation between $H_{sw0}^-$ and $\chi_{in}$

We now calculate the initial switching field of the particle along the *x* direction. This is done by requiring $F_{\alpha\alpha}(\pi/2,0)F_{\beta\beta}(\pi/2,0) - F_{\alpha\beta}^{2}(\pi/2,0) = 0$ and solving this equation for *H*. Using (1) and $E_{\alpha\beta}(\pi/2,0) = 0$, we obtain the result

$$H_{sw0^-} = [E_{\alpha\alpha}(\pi/2,0) + E_{\beta\beta}(\pi/2,0) - | E_{\alpha\alpha}(\pi/2,0) + E_{\beta\beta}(\pi/2,0) |]/2M_s$$

$$= \min\{E_{\alpha\alpha}(\pi/2,0), E_{\beta\beta}(\pi/2,0)\}/M_s, \qquad (6)$$

where $\min\{E_{\alpha\alpha}(\pi/2,0), E_{\beta\beta}(\pi/2,0)\}$ means the minimum of $E_{\alpha\alpha}(\pi/2,0)$ and $E_{\beta\beta}(\pi/2,0)$.

To find the correlation between $H_{sw0^-}$ and the initial susceptibility $\chi_{in}$, we calculate the maximal value of $\chi_{in}$. From (4), $\chi_{in}$ attains the maximum when $e^2_{Hy} = 1$ or $e^2_{Hz} = 1$, depending on the magnitude of $E_{\alpha\alpha}(\pi/2,0)$ and $E_{\beta\beta}(\pi/2,0)$. The maximum of $\chi_{in}$ is $\chi_{in,\max} = M_s^2 / \min\{E_{\alpha\alpha}(\pi/2,0), E_{\beta\beta}(\pi/2,0)\}$. $e^2_{Hy} = 1$ or $e^2_{Hz} = 1$ means that the external field **H** is either along the *y*-axis or the *z*-axis, which is perpendicular to the *x*-axis. This is consistent with the fact that **M** initially pointed to the positive *x* direction, since **H** perpendicular to the *x*-axis gives maximal torque. From (6) and the definition of the intrinsic pinning field $H_{p0} \equiv M_s / \chi_{in,\max}$ we immediately obtain $H_{p0} = H_{sw0^-}$.

The key to understand the complementary nature of coercivity enhancement and exchange bias in the F-AF exchange coupled systems is the coherence of magnetization.

The proof has been based on the coherent rotation of the F magnetization. Thus the complementary nature of coercivity enhancement and exchange bias confirms the rotation nature of the F magnetization in the F-AF systems. In fact, from the proof, the relationship $H_{sw0^-} = M_s / \chi_{in,\max}$ always holds for a single domain particle. In that sense, the complementary nature of coercivity enhancement and exchange bias in the F-AF exchange coupled systems is a special case for this relation. $H_{sw0^-}$ is the minimal reverse field to cause the magnetization to rotate. On the other hand, $\chi_{in,max}$ corresponds to the maximum rotation of magnetization under an (infinitesimally small) external field applied at some direction perpendicular to the magnetization. The external filed is not applied in the same direction in these two cases. However, the relation $H_{sw0^-} = M_s / \chi_{in,\max}$ implies the same physical process. In other words, for a reverse field $H = H_{sw0^-}$ the magnetization rotates in the direction that would attain the maximal initial susceptibility. When the external field is applied opposite to the magnetization, Zeeman energy has axial symmetry with respect to the magnetization direction. At $H = H_{sw0^-}$, the magnetization is no longer stable and rotates in the direction that requires minimal energy increase. On the other hand, when the rotation of magnetization costs minimal energy, it attains maximal initial susceptibility. This is evident from $\chi_{in,\max} = M_s^2 / \min\{E_{\alpha\alpha}(\pi/2,0), E_{\beta\beta}(\pi/2,0)\}$ and (5).

# IV. The difference between the initial switching field and the actual switching field

We distinguish between the initial switching field $H_{sw0}^-$ and the actual switching field. The initial switching field is defined as the reverse field at which the magnetization becomes unstable. In F-AF systems, sometimes it is found experimentally that the initial switching field is not the actual switching field for the F magnetization. That is to say, the F magnetization does not switch yet at $H = H_{sw0^-}$. Instead, it switches at a slightly larger field $H_{sw}^-$. Fig. 3(a) shows one such example. In the graph, $H_{sw0}^-$ is the field at which the hysteresis loop begins to "curve" from the horizontal line of positive saturation. Note the actual switching field, at which the F magnetization changes to the reverse direction abruptly, is larger than the initial switching field. Appendix B shows that this happens when the overall anisotropy energy is asymmetric with respect to the stable direction of the magnetization in the absence of external field. To study the switching behavior, we need to know the energy profile for $H \geq H_{sw0^-}$ and we take the layered F-AF system as an example. For simplicity, we assume that the overall anisotropy of the F layer includes only a unidirectional component and a uniaxial component. Other cases can be discussed similarly. The easy direction of the unidirectional component of the overall anisotropy (denoted by e.d., also called the pinned direction) is assumed to differ from the easy axis of the uniaxial component (denoted by e.a.), shown in Fig. 3(b). Suppose without any external field, the F magnetization initially stabilizes at some intermediate direction between e.d. and e.a., corresponding to an energy minimum state. In the figure, the external field is applied opposite to the stable direction. Thus the overall anisotropy of the F layer is not symmetric with respect to that stable direction and the energy barriers have different heights on the left and right hand sides of the F magnetization (see Fig. 3(c)). If a field **H** opposite to the stable direction with increasing magnitude is applied, the location of the minimum in the energy profile does not change for $H < H_{sw0^-}$. But for $H > H_{sw0^-}$, the above direction is no longer an energy minimum state. However, due to the asymmetry of the energy, a local minimum exists near that direction. Or equivalently the energy minimum shifts away from that direction continuously. This shift becomes larger as $H$ increases. In this particular example, the shift is always away from the e.d. direction. Consequently, the magnetization rotates away from that direction continuously. No switching occurs until at some stage when $H$ is so large that no energy minimum exists near that direction. When this happens, the magnetization will switch and the corresponding field is the actual switching field $H_{sw}^-$, which is greater than $H_{sw0}^-$. At $H = H_{sw0^-}$ the F magnetization is unstable. More precisely, the state is metastable. It corresponds to a kink point in the energy profile. It is clear that the asymmetry of the energy profile has a great impact on the magnetization behavior. This is quite different from the symmetric case, where the pinned direction is the stable direction under $H < H_{sw0^-}$, and for $H \geq H_{sw0^-}$ the pinned direction is unstable and corresponds to a maximal point in the energy profile. No energy minimum exists near the pinned direction for $H \geq H_{sw0^-}$. In that case $H_{sw}^-$ coincides with $H_{sw0}^-$, resulting in a square hysteresis loop. In the asymmetric case a loop curved between $H_{sw}^-$ and $H_{sw0}^-$ is observed, as shown in Fig. 3(a).

The above argument can be verified by considering the free energy of the F-AF layered system. To be consistent with the notations in the previous sections, let $\delta_1$ and $\delta_2$ be the angles of e.d. and e.a. and $\alpha$ be the angle of the F magnetization, all from the stable direction of F magnetization in the absence of external field. When an external field is applied opposite to the stable direction, the free energy density of the system is

$$F(\alpha) = M_s H \cos\alpha - K_{ed} \cos(\alpha - \delta_1) + K_{ea} \sin^2(\alpha - \delta_2)$$

where $K_{ed}$ and $K_{ea}$ denote the anisotropy constants for the unidirectional component and the uniaxial component, respectively. The derivatives of $F(\alpha)$ are

$$F'(\alpha) = -M_s H \sin\alpha + K_{ed} \sin(\alpha - \delta_1) + K_{ea} \sin 2(\alpha - \delta_2),$$

$$F''(\alpha) = -M_s H \cos\alpha + K_{ed} \cos(\alpha - \delta_1) + 2K_{ea} \cos 2(\alpha - \delta_2),$$

$$F'''(\alpha) = M_s H \sin\alpha - K_{ed} \sin(\alpha - \delta_1) - 4K_{ea} \sin 2(\alpha - \delta_2),$$

$$F^{(4)}(\alpha) = M_s H \cos\alpha - K_{ed} \cos(\alpha - \delta_1) - 8K_{ea} \cos 2(\alpha - \delta_2).$$

$\alpha = 0$ is the stable state for $H < H_{sw0^-}$. At $H = H_{sw0^-}$ the first and second derivatives of $F(\alpha)$ at $\alpha = 0$ vanish and the magnetization becomes unstable. Thus we have

$$F'(0) = -K_{ed} \sin\delta_1 - K_{ea} \sin 2\delta_2 = 0,$$

$$F''(0) = -M_s H_{sw0^-} + K_{ed} \cos\delta_1 + 2K_{ea} \cos 2\delta_2 = 0.$$

To study the energy profile near $\alpha = 0$ for $H = H_{sw0^-}$, we also need to calculate these higher order derivatives of $F(\alpha)$:

$$F'''(0) = K_{ed} \sin\delta_1 + 4K_{ea} \sin 2\delta_2$$

$$= -[-K_{ed}\sin\delta_1 - K_{ea}\sin 2\delta_2] + 3K_{ea}\sin 2\delta_2$$

$$= -F'(0) + 3K_{ea}\sin 2\delta_2$$

$$= 3K_{ea}\sin 2\delta_2,$$

$$F^{(4)}(0) = M_s H_{sw0^-} - K_{ed}\cos\delta_1 - 8K_{ea}\cos 2\delta_2$$

$$= -[-M_s H_{sw0^-} + K_{ed}\cos\delta_1 + 2K_{ea}\cos 2\delta_2] - 6K_{ea}\cos 2\delta_2$$

$$= -F''(0) - 6K_{ea}\cos 2\delta_2$$

$$= -6K_{ea}\cos 2\delta_2.$$

If $\delta_2 \neq 0$ we have $F'''(0) \neq 0$ and $\alpha = 0$ is a kink point of the free energy. Consequently, an energy minimum near $\alpha = 0$ exists for $H_{sw0^-} < H < H_{sw^-}$. On the other hand, if $\delta_2 = 0$ we have $F'''(0) = 0$ and $F^{(4)}(0) = -6K_{ea} < 0$. Thus $\alpha = 0$ is a maximum point of the free energy for $H \geq H_{sw0^-}$.

Fig. 3(c) shows schematically the changes of energy profile for the asymmetric case, especially for the reverse field near $H_{sw0}^-$ and $H_{sw}^-$. In the figure, $\alpha$ is the angle from the stable direction. Note the shift of the location of the energy minimum and the change of energy profile near $H_{sw0}^-$ and $H_{sw}^-$. It must be pointed out that both for symmetric and asymmetric cases, the relationship $H_{p0} = H_{sw0^-}$ holds strictly, as proved in section III.

## V. Conclusion

Complementary nature of coercivity enhancement and exchange bias is generalized from the layered ferro-antiferromagnet (F-AF) exchange coupled systems to F-AF systems with arbitrary configuration. It is proved based on the coherent rotation of the F magnetization. By absorbing the effect of F-AF coupling into the anisotropy of the F part, we are able to avoid the complicated nature of the F-AF coupling at the interface and keep the generality of the discussion. A general discussion for the initial susceptibility of a single domain particle is provided. The fundamental correlation between the maximal initial susceptibility and the initial switching field along an arbitrary easy direction of a single domain particle is analyzed. The difference between the initial switching field and the actual switching field along an arbitrary easy direction is also discussed.

## ACKNOWLEDGMENTS

We thank C. Papusoi in the MINT Center, T. Z. Mai and D. Halpern in the Department of Mathematics at the University of Alabama for helpful discussions. This project was funded by grants NSF-DMS-0207137 & NSF-DMR-0213985.

## Appendix A

Now we give a mathematical proof of the claim that $E_{\alpha\beta}(\pi/2,0)$ vanishes for some appropriate choice of the $y$ and $z$ axes when the positive $x$ direction is an easy direction. Without loss of generality, suppose $E_{\alpha\beta}(\pi/2,0) \neq 0$ in the current coordinate system. After rotating the $y$ and $z$ axes in the $y$-$z$ plane with an angle $\gamma$, we obtain a new coordinate system $xy'z'$ (see Fig. 4). The $y'$ and $z'$ components of the magnetization are denoted by $M_{y'}$ and $M_{z'}$. Thus we have

$$M_y = \cos\gamma M_{y'} + \sin\gamma M_{z'}$$

$$M_z = -\sin\gamma M_{y'} + \cos\gamma M_{z'}.$$

The $y$ and $z$ components of the magnetization $M_y$ and $M_z$ are expressed in terms of $\alpha$ and $\beta$ by $M_y = M_s \sin\alpha \sin\beta$ and $M_z = M_s \cos\alpha$. From that we can obtain

$$\sin\alpha \sin\beta = \cos\gamma \sin\alpha' \sin\beta' + \sin\gamma \cos\alpha',$$

$$\cos\alpha = -\sin\gamma \sin\alpha' \sin\beta' + \cos\gamma \cos\alpha',$$

where $\alpha'$ and $\beta'$ are the polar and azimuthal angles of the magnetization in the new coordinate system.

The above two relations give the dependences of $\alpha$ and $\beta$ on $\alpha'$ and $\beta'$. Taking partial derivatives with respect to $\alpha'$ and $\beta'$ on both sides of these equations yields

$$\cos\alpha\sin\beta\partial\alpha/\partial\alpha'+\sin\alpha\cos\beta\partial\beta/\partial\alpha'=\cos\gamma\cos\alpha'\sin\beta'-\sin\gamma\sin\alpha',$$

$$\cos\alpha\sin\beta\partial\alpha/\partial\beta'+\sin\alpha\cos\beta\partial\beta/\partial\beta'=\cos\gamma\sin\alpha'\cos\beta',$$

$$-\sin\alpha\partial\alpha/\partial\alpha'=-\sin\gamma\cos\alpha'\sin\beta'-\cos\gamma\sin\alpha',$$

$$-\sin\alpha\partial\alpha/\partial\beta'=-\sin\gamma\sin\alpha'\cos\beta'.$$

Note after the rotation the x-axis remains unchanged. For the positive $x$ direction, $\alpha=\pi/2, \beta=0$ and hence $\alpha'=\pi/2, \beta'=0$. Thus we have

$$\partial\beta/\partial\alpha'|_{(\pi/2,0)}=-\sin\gamma, \tag{A1}$$

$$\partial\beta/\partial\beta'|_{(\pi/2,0)}=\cos\gamma, \tag{A2}$$

$$-\partial\alpha/\partial\alpha'|_{(\pi/2,0)}=-\cos\gamma \text{ or } \partial\alpha/\partial\alpha'|_{(\pi/2,0)}=\cos\gamma, \tag{A3}$$

$$-\partial\alpha/\partial\beta'|_{(\pi/2,0)}=-\sin\gamma \text{ or } \partial\alpha/\partial\beta'|_{(\pi/2,0)}=\sin\gamma. \tag{A4}$$

On the other hand

$$E_{\alpha'}\equiv\partial E/\partial\alpha'=\partial E/\partial\alpha\partial\alpha/\partial\alpha'+\partial E/\partial\beta\partial\beta/\partial\alpha'=E_\alpha\partial\alpha/\partial\alpha'+E_\beta\partial\beta/\partial\alpha',$$

and

$$E_{\alpha'\beta'}\equiv\partial^2 E/\partial\alpha'\partial\beta'=(E_{\alpha\alpha}\partial\alpha/\partial\beta'+E_{\alpha\beta}\partial\beta/\partial\beta')\partial\alpha/\partial\alpha'+E_\alpha\partial^2\alpha/\partial\alpha'\partial\beta'$$

$$+(E_{\beta\alpha}\partial\alpha/\partial\beta'+E_{\beta\beta}\partial\beta/\partial\beta')\partial\beta/\partial\alpha'+E_\beta\partial^2\beta/\partial\alpha'\partial\beta'.$$

Since the x-axis is an easy axis, $E_\alpha(\pi/2,0)=0$, $E_\beta(\pi/2,0)=0$. Assuming $E_{\alpha\beta}=E_{\beta\alpha}$ and using (A1) – (A4), we obtain

$$E_{\alpha'\beta'}(\pi/2,0)=[\sin\gamma E_{\alpha\alpha}(\pi/2,0)+\cos\gamma E_{\alpha\beta}(\pi/2,0)]\cos\gamma$$

$$+[\sin\gamma E_{\alpha\beta}(\pi/2,0)+\cos\gamma E_{\beta\beta}(\pi/2,0)](-\sin\gamma)$$

$$= \sin\gamma\cos\gamma[E_{\alpha\alpha}(\pi/2,0) - E_{\beta\beta}(\pi/2,0)]$$

$$+ (\cos^2\gamma - \sin^2\gamma)E_{\alpha\beta}(\pi/2,0)$$

$$= \sin 2\gamma[E_{\alpha\alpha}(\pi/2,0) - E_{\beta\beta}(\pi/2,0)] + \cos 2\gamma E_{\alpha\beta}(\pi/2,0).$$

We choose $\gamma$ such that

$$\cot 2\gamma = \frac{E_{\beta\beta}(\pi/2,0) - E_{\alpha\alpha}(\pi/2,0)}{2E_{\alpha\beta}(\pi/2,0)}.$$

Then $E_{\alpha'\beta'}(\pi/2,0) = 0$.

# Appendix B

In this appendix, we derive a necessary and sufficient condition for $H_{sw^-} \neq H_{sw0^-}$ for a single domain particle and discuss a method to determine $H_{sw0^-}$ from the hysteresis loop. The third partial derivatives of the overall anisotropy are assumed to exist in a neighborhood of the initial F magnetization direction.

(1) **Necessary and sufficient condition for $H_{sw^-} \neq H_{sw0^-}$ for a single domain particle**

$H_{sw0^-}$ has been defined as the reverse field along an easy axis at which the magnetization along the easy axis becomes unstable. By "unstable", it does not necessarily mean that the magnetization will switch immediately. The magnetization direction can still undergo smooth changes afterwards. Whether it switches or not depends on the energy profile. Thus we need to examine the energy profile and find the local minima for the system at $H \geq H_{sw0^-}$, from which the condition for $H_{sw^-} \neq H_{sw0^-}$ can be obtained.

As before, let $\alpha$ and $\beta$ be the polar and azimuthal angles for the magnetization. Suppose that the $+x$ direction, i.e., $\alpha = \pi/2$ and $\beta = 0$, is an easy direction of the magnetization and $H$ is applied along the $-x$ direction. The magnetization stays in the $+x$ direction when $H < H_{sw0^-}$. For $H > H_{sw0^-}$, the magnetization becomes unstable. Our task is to find the stable magnetization direction when $H > H_{sw0^-}$. For that, we increase an infinitesimal amount of the field $\delta H > 0$ from $H_{sw0^-}$ and find the condition for a stable state to exist in a neighborhood of the $+x$ direction. With $H = H_{sw0^-} + \delta H$ we seek solutions to

$$F_\alpha \equiv \partial F / \partial \alpha = 0, \tag{B1}$$

$$F_\beta \equiv \partial F / \partial \beta = 0. \tag{B2}$$

Under what conditions can we find a stable solution in a neighborhood of the $+x$ direction?

We discuss a necessary condition for that first. Note $F_\alpha = E_\alpha + M_s H \cos\alpha \cos\beta$ and $F_\beta = E_\beta - M_s H \sin\alpha \sin\beta$. For $|\delta\alpha| \ll 1$ and $|\delta\beta| \ll 1$,

$$F_\alpha(\pi/2+\delta\alpha,\delta\beta) = \sum_{v=0}^{n} \frac{([\delta\alpha\frac{\partial}{\partial\alpha}+\delta\beta\frac{\partial}{\partial\beta}]^v F_\alpha)(\pi/2,0)}{v!}$$

$$+ \frac{([\delta\alpha\frac{\partial}{\partial\alpha}+\delta\beta\frac{\partial}{\partial\beta}]^{n+1} F_\alpha)(\pi/2+\eta\delta\alpha,\eta\delta\beta)}{(n+1)!}, \qquad (B3)$$

where $\eta \in [0,1]$.

Suppose that the coordinate system has been selected properly so that $E_{\alpha\beta}(\pi/2,0) = E_{\beta\alpha}(\pi/2,0) = 0$. Without loss of generality we assume $E_{\alpha\alpha}(\pi/2,0) \leq E_{\beta\beta}(\pi/2,0)$. From our result in section III,

$$H_{sw0^-} = \min\{E_{\alpha\alpha}(\pi/2,0), E_{\beta\beta}(\pi/2,0)\}/M_s = E_{\alpha\alpha}(\pi/2,0)/M_s. \qquad (B4)$$

From (B1), (B3), and (B4), we obtain for $\delta H > 0$,

$$F_\alpha(\pi/2+\delta\alpha,\delta\beta) = F_\alpha(\pi/2,0) + ([\delta\alpha\frac{\partial}{\partial\alpha}+\delta\beta\frac{\partial}{\partial\beta}]F_\alpha)(\pi/2,0)$$

$$+ \frac{([\delta\alpha\frac{\partial}{\partial\alpha}+\delta\beta\frac{\partial}{\partial\beta}]^2 F_\alpha)(\pi/2,0)}{2!}$$

$$+ \frac{([\delta\alpha\frac{\partial}{\partial\alpha}+\delta\beta\frac{\partial}{\partial\beta}]^3 F_\alpha)(\pi/2+\eta_1\delta\alpha,\eta_1\delta\beta)}{3!}$$

$$= E_\alpha(\pi/2,0) + [E_{\alpha\alpha}(\pi/2,0) - M_s H]\delta\alpha + E_{\alpha\beta}(\pi/2,0)\delta\beta$$

$$+ \frac{1}{2}[E_{\alpha\alpha\alpha}(\pi/2,0)(\delta\alpha)^2 + 2E_{\alpha\alpha\beta}(\pi/2,0)(\delta\alpha)(\delta\beta)$$

$$+ E_{\alpha\beta\beta}(\pi/2,0)(\delta\beta)^2]$$

$$+ \frac{([\delta\alpha \frac{\partial}{\partial\alpha} + \delta\beta \frac{\partial}{\partial\beta}]^3 F_\alpha)(\pi/2 + \eta_1\delta\alpha, \eta_1\delta\beta)}{3!}$$

$$= E_\alpha(\pi/2,0) + [E_{\alpha\alpha}(\pi/2,0) - M_s(H_{sw0-} + \delta H)]\delta\alpha$$

$$+ E_{\alpha\beta}(\pi/2,0)\delta\beta + \frac{1}{2}[E_{\alpha\alpha\alpha}(\pi/2,0)(\delta\alpha)^2$$

$$+ 2E_{\alpha\alpha\beta}(\pi/2,0)(\delta\alpha)(\delta\beta) + E_{\alpha\beta\beta}(\pi/2,0)(\delta\beta)^2]$$

$$+ \frac{([\delta\alpha \frac{\partial}{\partial\alpha} + \delta\beta \frac{\partial}{\partial\beta}]^3 F_\alpha)(\pi/2 + \eta_1\delta\alpha, \eta_1\delta\beta)}{3!}$$

$$= \frac{1}{2}[E_{\alpha\alpha\alpha}(\pi/2,0)(\delta\alpha)^2 + 2E_{\alpha\alpha\beta}(\pi/2,0)(\delta\alpha)(\delta\beta)$$

$$+ E_{\alpha\beta\beta}(\pi/2,0)(\delta\beta)^2] - M_s(\delta H)(\delta\alpha)$$

$$+ \frac{([\delta\alpha \frac{\partial}{\partial\alpha} + \delta\beta \frac{\partial}{\partial\beta}]^3 F_\alpha)(\pi/2 + \eta_1\delta\alpha, \eta_1\delta\beta)}{3!}$$

$$= 0, \tag{B5}$$

where $\eta_1 \in [0,1]$. In (B5), $E_\alpha(\pi/2,0) = 0$ since the +x direction is an easy direction. Similarly from (B2) we obtain

$$F_\beta(\pi/2 + \delta\alpha, \delta\beta) = F_\beta(\pi/2,0) + ([\delta\alpha \frac{\partial}{\partial\alpha} + \delta\beta \frac{\partial}{\partial\beta}]F_\beta)(\pi/2,0)$$

$$+ \frac{1}{2}([\delta\alpha \frac{\partial}{\partial\alpha} + \delta\beta \frac{\partial}{\partial\beta}]^2 F_\beta)(\pi/2 + \eta_2\delta\alpha, \eta_2\delta\beta)$$

$$= F_\beta(\pi/2,0) + F_{\beta\alpha}(\pi/2,0)\delta\alpha + F_{\beta\beta}(\pi/2,0)\delta\beta$$

$$+ \frac{1}{2}([\delta\alpha \frac{\partial}{\partial\alpha} + \delta\beta \frac{\partial}{\partial\beta}]^2 F_\beta)(\pi/2 + \eta_2\delta\alpha, \eta_2\delta\beta)$$

$$= E_\beta(\pi/2,0) + E_{\beta\alpha}(\pi/2,0)\delta\alpha + [E_{\beta\beta}(\pi/2,0) - M_s H]\delta\beta$$

$$+\frac{1}{2}([\delta\alpha\frac{\partial}{\partial\alpha}+\delta\beta\frac{\partial}{\partial\beta}]^2 F_\beta)(\pi/2+\eta_2\delta\alpha,\eta_2\delta\beta)$$

$$=[E_{\beta\beta}(\pi/2,0)-M_s(H_{sw0-}+\delta H)]\delta\beta$$

$$+\frac{1}{2}([\delta\alpha\frac{\partial}{\partial\alpha}+\delta\beta\frac{\partial}{\partial\beta}]^2 F_\beta)(\pi/2+\eta_2\delta\alpha,\eta_2\delta\beta)$$

$$=[E_{\beta\beta}(\pi/2,0)-E_{\alpha\alpha}(\pi/2,0)-M_s\delta H]\delta\beta$$

$$+\frac{1}{2}([\delta\alpha\frac{\partial}{\partial\alpha}+\delta\beta\frac{\partial}{\partial\beta}]^2 F_\beta)(\pi/2+\eta_2\delta\alpha,\eta_2\delta\beta)$$

$$=0 \qquad (B6)$$

where $\eta_2 \in [0,1]$. Again $E_\beta(\pi/2,0)=0$ since the +x direction is an easy direction.

Depending on the magnitudes of $E_{\alpha\alpha}(\pi/2,0)$ and $E_{\beta\beta}(\pi/2,0)$ we consider two different cases:

(i) $E_{\alpha\alpha}(\pi/2,0) < E_{\beta\beta}(\pi/2,0)$

In this case we obtain from (B6)

$$\delta\beta = \frac{([\delta\alpha\frac{\partial}{\partial\alpha}+\delta\beta\frac{\partial}{\partial\beta}]^2 F_\beta)(\pi/2+\eta_2\delta\alpha,\eta_2\delta\beta)}{2[E_{\beta\beta}(\pi/2,0)-E_{\alpha\alpha}(\pi/2,0)-M_s\delta H]}.$$

Or equivalently

$$\frac{\delta \beta}{\delta \alpha}\{1-\frac{F_{\beta\beta\beta}(\pi/2+\eta_2\delta\alpha,\eta_2\delta\beta)}{2[E_{\beta\beta}(\pi/2,0)-E_{\alpha\alpha}(\pi/2,0)-M_s\delta H]}\delta\beta\}$$

$$=\frac{1}{2[E_{\beta\beta}(\pi/2,0)-E_{\alpha\alpha}(\pi/2,0)-M_s\delta H]}[F_{\beta\alpha\alpha}(\pi/2+\eta_2\delta\alpha,\eta_2\delta\beta)\delta\alpha$$

$$+2F_{\beta\alpha\beta}(\pi/2+\eta_2\delta\alpha,\eta_2\delta\beta)\delta\beta].$$

Taking the limit as $\delta H \to 0$ (for necessary condition, $\delta\alpha \to 0$, $\delta\beta \to 0$), we obtain

$$\delta\beta/\delta\alpha = 0.$$

Therefore in the limit as $\delta H \to 0$, (B5) is simplified to

$$E_{\alpha\alpha\alpha}(\pi/2,0) - 2M_s \delta H/\delta\alpha = 0,$$

or

$$E_{\alpha\alpha\alpha}(\pi/2,0)\delta\alpha/\delta H - 2M_s = 0.$$

In order for $\delta\alpha/\delta H < \infty$, we need $E_{\alpha\alpha\alpha}(\pi/2,0) \neq 0$, and we obtain

$$\delta\alpha/\delta H = 2M_s/E_{\alpha\alpha\alpha}(\pi/2,0). \tag{B7}$$

Thus $E_{\alpha\alpha\alpha}(\pi/2,0) \neq 0$ is a necessary condition for (B5) to have solutions, i.e., a necessary condition for a stable sate to exist at the neighborhood of the +x direction.

In fact, $E_{\alpha\alpha\alpha}(\pi/2,0) \neq 0$ is also the sufficient condition. To see it, notice when $\delta H \to 0$,

$$F_{\alpha\alpha}(\pi/2+\delta\alpha,\delta\beta) = E_{\alpha\alpha}(\pi/2,0) + E_{\alpha\alpha\alpha}(\pi/2,0)\delta\alpha + E_{\alpha\alpha\beta}(\pi/2,0)\delta\beta$$

$$-M_s H + \frac{1}{2}([\delta\alpha\frac{\partial}{\partial\alpha}+\delta\beta\frac{\partial}{\partial\beta}]^2 F_{\alpha\alpha})(\pi/2+\eta_3\delta\alpha,\eta_3\delta\beta)$$

$$= E_{\alpha\alpha}(\pi/2,0) + E_{\alpha\alpha\alpha}(\pi/2,0)\delta\alpha + E_{\alpha\alpha\beta}(\pi/2,0)\delta\beta$$

$$-M_s(H_{sw0-} + \delta H)$$

$$+\frac{1}{2}([\delta\alpha\frac{\partial}{\partial\alpha} + \delta\beta\frac{\partial}{\partial\beta}]^2 F_{\alpha\alpha})(\pi/2 + \eta_3\delta\alpha, \eta_3\delta\beta)$$

$$= E_{\alpha\alpha\alpha}(\pi/2, 0)\delta\alpha + E_{\alpha\alpha\beta}(\pi/2, 0)\delta\beta - M_s\delta H$$

$$+\frac{1}{2}([\delta\alpha\frac{\partial}{\partial\alpha} + \delta\beta\frac{\partial}{\partial\beta}]^2 F_{\alpha\alpha})(\pi/2 + \eta_3\delta\alpha, \eta_3\delta\beta)$$

$$\sim M_s\delta H$$

where $\eta_3 \in [0,1]$. In the above derivation, (B4) and (B7) were used.

Similarly we have

$$F_{\alpha\beta}(\pi/2 + \delta\alpha, \delta\beta) = E_{\alpha\beta}(\pi/2, 0) + E_{\alpha\beta\alpha}(\pi/2, 0)\delta\alpha + E_{\alpha\beta\beta}(\pi/2, 0)\delta\beta$$

$$+\frac{1}{2}([\delta\alpha\frac{\partial}{\partial\alpha} + \delta\beta\frac{\partial}{\partial\beta}]^2 F_{\alpha\alpha})(\pi/2 + \eta_4\delta\alpha, \eta_4\delta\beta)$$

$$= E_{\alpha\beta\alpha}(\pi/2, 0)\delta\alpha + E_{\alpha\beta\beta}(\pi/2, 0)\delta\beta$$

$$+\frac{1}{2}([\delta\alpha\frac{\partial}{\partial\alpha} + \delta\beta\frac{\partial}{\partial\beta}]^2 F_{\alpha\alpha})(\pi/2 + \eta_4\delta\alpha, \eta_4\delta\beta)$$

$$\sim E_{\alpha\beta\alpha}(\pi/2, 0)\delta\alpha,$$

where $\eta_4 \in [0,1]$, and

$$F_{\beta\beta}(\pi/2 + \delta\alpha, \delta\beta) = E_{\beta\beta}(\pi/2, 0) - M_s H$$

$$+\frac{1}{2}([\delta\alpha\frac{\partial}{\partial\alpha} + \delta\beta\frac{\partial}{\partial\beta}]^2 F_{\alpha\alpha})(\pi/2 + \eta_5\delta\alpha, \eta_5\delta\beta)$$

$$= E_{\beta\beta}(\pi/2, 0) - M_s(H_{sw0-} + \delta H)$$

$$+ \frac{1}{2}([\delta\alpha \frac{\partial}{\partial\alpha} + \delta\beta \frac{\partial}{\partial\beta}]^2 F_{\alpha\alpha})(\pi/2 + \eta_5\delta\alpha, \eta_5\delta\beta)$$

$$= E_{\beta\beta}(\pi/2,0) - E_{\alpha\alpha}(\pi/2,0) - M_s\delta H$$

$$+ \frac{1}{2}([\delta\alpha \frac{\partial}{\partial\alpha} + \delta\beta \frac{\partial}{\partial\beta}]^2 F_{\alpha\alpha})(\pi/2 + \eta_5\delta\alpha, \eta_5\delta\beta)$$

$$\sim E_{\beta\beta}(\pi/2,0) - E_{\alpha\alpha}(\pi/2,0),$$

where $\eta_5 \in [0,1]$.

To check the stability of the state $\alpha = \pi/2 + \delta\alpha$, $\beta = \delta\beta$, we consider the sign of $F_{\alpha\alpha}(\pi/2 + \delta\alpha, \delta\beta)F_{\beta\beta}(\pi/2 + \delta\alpha, \delta\beta) - [F_{\alpha\beta}(\pi/2 + \delta\alpha, \delta\beta)]^2$ as $\delta H \to 0$.

$$F_{\alpha\alpha}(\pi/2 + \delta\alpha, \delta\beta)F_{\beta\beta}(\pi/2 + \delta\alpha, \delta\beta) - [F_{\alpha\beta}(\pi/2 + \delta\alpha, \delta\beta)]^2$$

$$\sim [E_{\beta\beta}(\pi/2,0) - E_{\alpha\alpha}(\pi/2,0)]M_s\delta H - [E_{\alpha\beta\alpha}(\pi/2,0)\delta\alpha]^2$$

$$\sim [E_{\beta\beta}(\pi/2,0) - E_{\alpha\alpha}(\pi/2,0)]M_s\delta H > 0$$

since $\delta H > 0$ and $E_{\alpha\alpha}(\pi/2,0) < E_{\beta\beta}(\pi/2,0)$ from the assumptions. The second order terms were dropped.

Therefore $\alpha = \pi/2 + \delta\alpha$, $\beta = \delta\beta$, is a stable direction for the magnetization at $H = H_{sw0^-} + \delta H$ and we proved that $E_{\alpha\alpha\alpha}(\pi/2,0) \neq 0$ is the necessary and sufficient condition for it. The case of $E_{\alpha\alpha}(\pi/2,0) > E_{\beta\beta}(\pi/2,0)$ can be discussed similarly and in that case, $E_{\beta\beta\beta}(\pi/2,0) \neq 0$ is the necessary and sufficient condition for the existence of a stable direction for the magnetization at $H = H_{sw0^-} + \delta H$.

$E_{\alpha\alpha\alpha}(\pi/2,0) \neq 0$ means that the anisotropy energy is asymmetric with respect to the *xz*-plane and $E_{\beta\beta\beta}(\pi/2,0) \neq 0$ means that the anisotropy energy is asymmetric with respect to the *xy*-plane. In the case of layered F-AF systems, the above conditions reduce to the anisotropy energy asymmetric with respect to the stable direction of the F magnetization in the absence of external field.

(ii) $E_{\alpha\alpha}(\pi/2,0) = E_{\beta\beta}(\pi/2,0)$

This case is more complicated. In this case the first order terms in (B6) vanish and we must keep all the second order terms. Hence, we need to solve two quadratic equations (B5) and (B6) jointly to obtain *δα/δH* and *δβ/δH*. The existence and the nature of solutions depend on the third derivatives of anisotropy energy *E* (with respect to *α* and *β*). Furthermore, even if a solution exists, whether or not it is a stable direction of the magnetization for $H = H_{sw0^-} + \delta H$ is still under questioning. We still need to use the criterion $F_{\alpha\alpha}(\pi/2+\delta\alpha,\delta\beta)F_{\beta\beta}(\pi/2+\delta\alpha,\delta\beta) - [F_{\alpha\beta}(\pi/2+\delta\alpha,\delta\beta)]^2 > 0$ to determine its stability, which again turns out to depend on the third derivatives of *E* (with respect to

$\alpha$ and $\beta$). Thus, in general no simple specific conclusion can be drawn for this case. Since in reality, the condition $E_{\alpha\alpha}(\pi/2,0) = E_{\beta\beta}(\pi/2,0)$ is seldom satisfied, we do not provide further detailed discussion here.

**(2) Methodology to obtain $H_{sw0}^-$ when it is not the switching field**

This question has significance in practice when $H_{sw^-} \neq H_{sw0^-}$ and we want to obtain $H_{sw0}^-$ accurately in experiments. We discuss the dependence of $M_x$, the component of the magnetization in the $+x$ direction, on $H = H_{sw0^-} + \Delta H$ when $\Delta H$ is small (recall that $H$ is applied in the $-x$ direction). For a general 3-d F-AF system, the physical processes involved in this discussion are assumed to be reversible. For a 2-d F-AF system, this condition is not required.

Note when $\Delta H$ is small, the magnetization direction will be close to the $+x$ direction.

$$M_x = M_s \sin\alpha \cos\beta$$
$$= M_s \sin(\pi/2 + \Delta\alpha)\cos(\Delta\beta) = M_s \cos(\Delta\alpha)\cos(\Delta\beta)$$
$$\approx M_s [1 - \frac{1}{2}(\Delta\alpha)^2][1 - \frac{1}{2}(\Delta\beta)^2]. \tag{B8}$$

Assume $E_{\alpha\alpha}(\pi/2,0) < E_{\beta\beta}(\pi/2,0)$. From the previous section we have $\Delta\beta/\Delta\alpha \approx 0$.

Thus from $F_\alpha(\pi/2+\Delta\alpha,0) = 0$ we get

$$E_\alpha(\pi/2+\Delta\alpha,0) + M_s H \cos(\pi/2+\Delta\alpha)$$

$$\approx E_\alpha(\pi/2,0) + E_{\alpha\alpha}(\pi/2,0)\Delta\alpha + \frac{1}{2}E_{\alpha\alpha\alpha}(\pi/2,0)(\Delta\alpha)^2$$

$$- M_s(H_{sw0-} + \Delta H)\sin(\Delta\alpha)$$

$$\approx E_{\alpha\alpha}(\pi/2,0)\Delta\alpha + \frac{1}{2}E_{\alpha\alpha\alpha}(\pi/2,0)(\Delta\alpha)^2 - M_s(H_{sw0-} + \Delta H)(\Delta\alpha)$$

$$= \frac{1}{2}E_{\alpha\alpha\alpha}(\pi/2,0)(\Delta\alpha)^2 - M_s(\Delta H)(\Delta\alpha)$$

$$= 0.$$

which yields

$$\Delta\alpha \approx \frac{2M_s}{E_{\alpha\alpha\alpha}(\pi/2,0)}\Delta H.$$

Using (B8) we obtain

$$\Delta M_x \equiv M_x - M_s \approx -\frac{1}{2}M_s(\Delta\alpha)^2 \approx -\frac{2M_s^3}{[E_{\alpha\alpha\alpha}(\pi/2,0)]^2}(\Delta H)^2.$$

Thus $H_{sw0}^-$ can be obtained by using the above relation for the part of $H \geq H_{sw0-}$ within a small range (see Fig. 3(a)).

FIG. 1. External field H and magnetization M in spherical coordinates. M initially pointed to the positive *x* direction, an easy direction of M.

FIG. 2. Plot of initial susceptibility of a single domain particle with $E_{\alpha\alpha}(\pi/2,0)/E_{\beta\beta}(\pi/2,0) = 1/2$, in the unit of $M_s^2/E_{\beta\beta}(\pi/2,0)$. The radii represent the initial susceptibility values in different directions. In the plot, the coordinates for the *x* axis and the *y* axis are from -1 to 1 and the coordinates for the *z* axis are from -2 to 2. The magnetization initially pointed to the positive *x* direction, an easy direction of the magnetization.

FIG. 3(a). Hysteresis loop along the F magnetization stable direction in the absence of external field, of a $Ni_{80}Fe_{20}$/FeMn F-AF coupled bilayer, measured by MOKE. (b). Schematic view of the easy direction of the unidirectional component e.d. (pinned direction), the easy axis of the uniaxial component e.a. of the overall anisotropy of the F layer, and the stable direction of F for an F-AF layered system in the absence of external field. (c) Schematic view of energy profiles for an F-AF layered system with asymmetry in the overall anisotropy, especially for the external field near $H_{sw0}^-$ and $H_{sw}^-$.

FIG. 4. Rotation of the coordinate system about the *x*-axis by an angle $\gamma$ giving a new coordinate system $xy'z'$.

[1] W. H. Meiklejohn and C. P. Bean, Phys. Rev. **102**, 1413 (1956).

[2] S. Riedling, M. Bauer, C. Mathieu, B. Hillbrands, R. Jungblut, J. Kohlhepp, and A. Reinders, J. Appl. Phys. **85**, 6648 (1999).

[3] C. Liu, J. Du, and G. J. Mankey, J. Vac. Sci. Technol. A **19**, 1213 (2001).

[4] H. Fujiwara, C. Hou, M. Sun, and H. S. Cho, IEEE Trans. Magn. **35**, 3082 (1999).

[5] H. S. Cho and H. Fujiwara, IEEE Trans. Magn. **35**, 3868 (1999).

[6] Chih-Huang Lai, Hideo Matsuyama, Robert L. White, Thomas C. Anthony, and Gary G. Bush, J. Appl. Phys. **79**, 6389 (1996).

[7] Haiwen Xi and Robert M. White, J. Appl. Phys. **86**, 5169 (1999).

[8] C. Hou, Ph. D. Dissertation Theis, University of Alabama, Physics Dept. (1999).

[9] C. Hou, H. Fujiwara, F. Ueda, H.S. Cho, MRS Proceedings **517**, 37 (1998).

[10] Z. Qian, J. M. Siversten, and J. H. Judy, J. Appl. Phys. **83**, 6825 (1998).

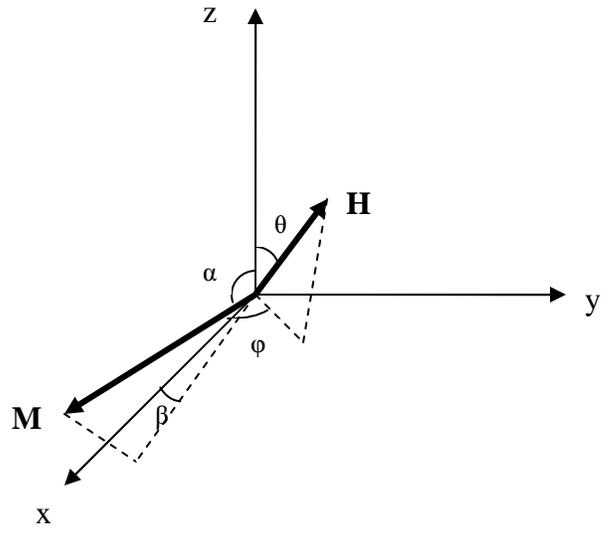

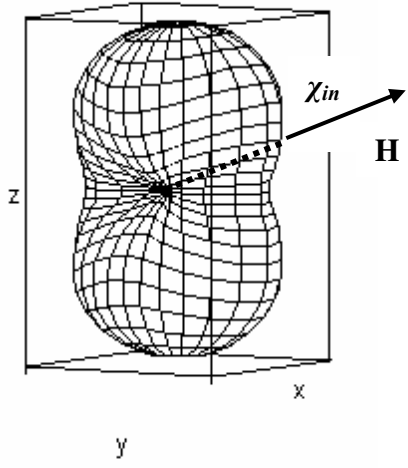

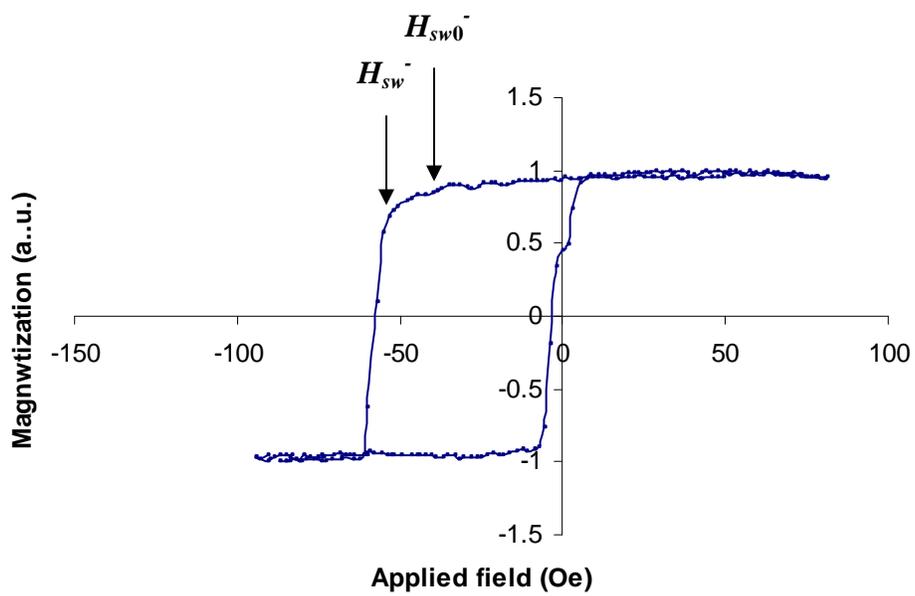

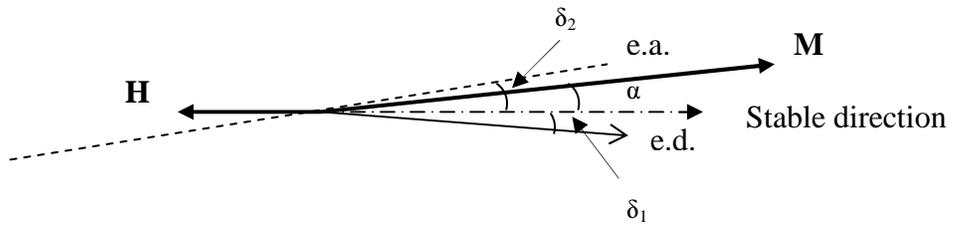

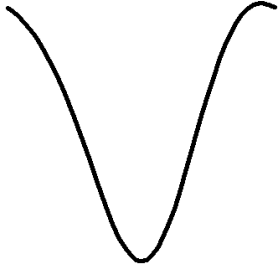 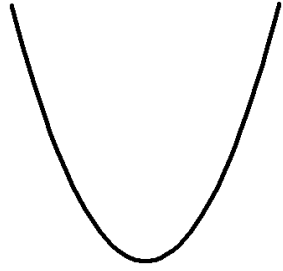

α from −π to π        enlarged: α from −0.05 to 0.05, minimum at $\alpha = 0$

$H = 0$

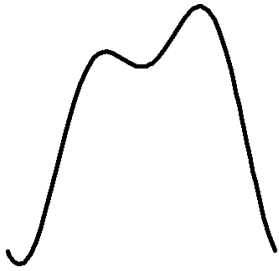 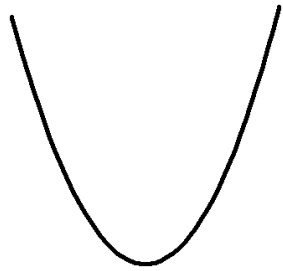

α from −π to π        enlarged: α from −0.05 to 0.05, minimum at $\alpha = 0$

$H < H_{sw0^-}$

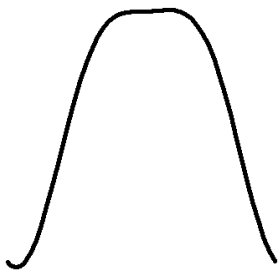 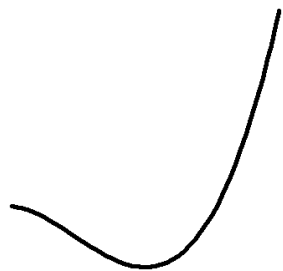

α from −π to π        enlarged: α from −0.05 to 0.05, minimum at α = 0

$H < H_{sw0^-}$, $H$ very close to $H_{sw0^-}$

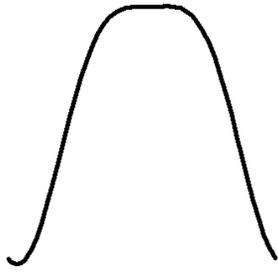 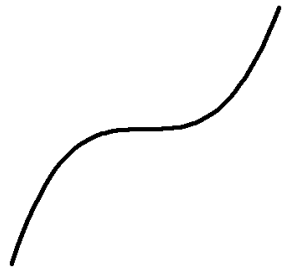

α from –π to π   enlarged: α from –0.05 to 0.05, kink point at α = 0

$$H = H_{sw0^-}$$

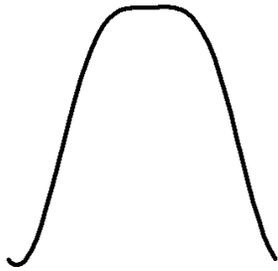 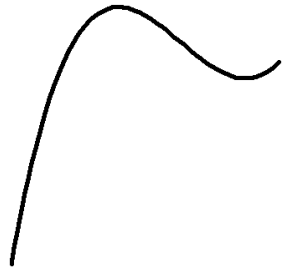

α from –π to π   enlarged: α from –0.04 to 0.06, minimum at some α > 0

$$H > H_{sw0^-}, H \text{ very close to } H_{sw0}^-$$

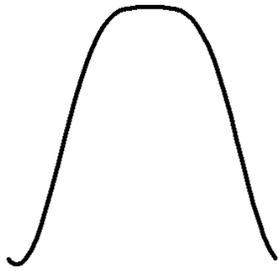 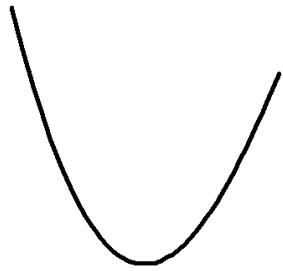

α from –π to π   enlarged: α from 0.2 to 0.3, minimum at some α > 0

$$H < H_{sw^-}, H \text{ very close to } H_{sw}^-$$

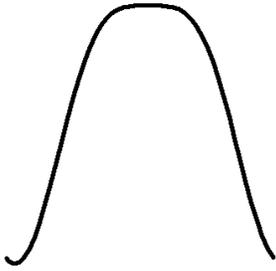 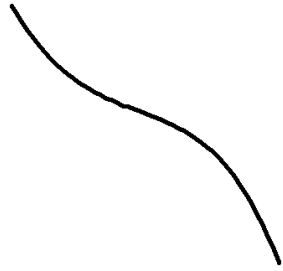

α from −π to π    enlarged: α from 0.26 to 0.36, no minimum exists

$H > H_{sw^-}$, $H$ very close to $H_{sw}^-$

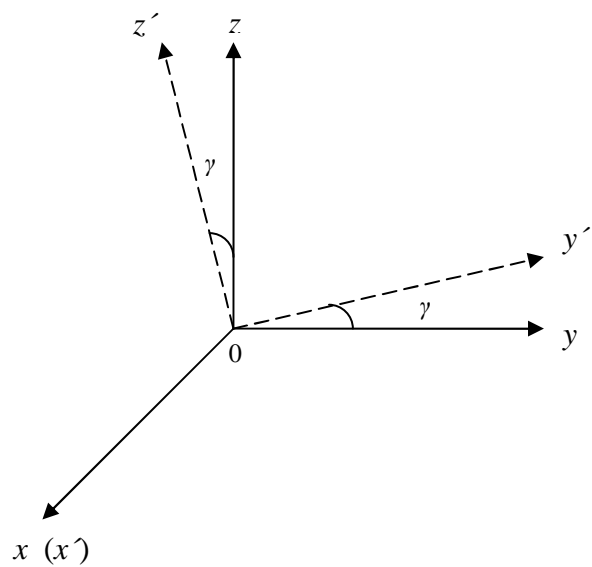